\documentstyle[preprint,aps,eqsecnum,psfig]{revtex}
\preprint{hep-th/0210272, INJE-TP-02-05}
\begin{document}
\tightenlines

\title{Holography in a Radiation-dominated  Universe with a
Positive Cosmological Constant}

\author{Rong-Gen Cai\footnote{e-mail address:
cairg@itp.ac.cn}}
\address{Institute of Theoretical Physics, Chinese
Academy of Sciences,\\
 P.O. Box 2735, Beijing 100080, China }
\author{Yun Soo Myung\footnote{e-mail
 address: ysmyung@physics.inje.ac.kr}}
 \address{ Relativity Research Center and School of Computer
Aided Science, \\
Inje University, Gimhae 621-749, Korea}

\maketitle

\begin{abstract}
We discuss the holographic principle in a radiation-dominated,
closed Friedmann-Robertson-Walker (FRW) universe with a positive
cosmological constant. By introducing a cosmological D-bound on
the  entropy of matter in the universe, we can write the Friedmann
equation governing the evolution of the universe in the form of
the Cardy formula. When the cosmological D-bound is saturated, the
Friedmann equation coincides with the Cardy-Verlinde formula
describing the entropy of radiation in the universe.  As a
concrete model, we consider a brane universe  in the background of
Schwarzschild-de Sitter black holes. It is found that the
cosmological D-bound is saturated when the brane crosses the black
hole horizon of the background. At that moment, the Friedmann
equation coincides with the Cardy-Verlinde formula describing the
entropy of radiation matter on the brane.

\end{abstract}

\newpage

\section{Introduction}

The holographic principle is perhaps one of fundamental principles
of nature, which relates a theory with gravity in $D$ dimensions
to a theory without gravity in lower dimensions~\cite{Hooft}.
Although we do not yet completely understand how the hologram of
gravity is realized, some beautiful examples have been found
through the AdS/CFT correspondence~\cite{AdS}.

In a seminal paper~\cite{Verl},  E. Verlinde found a quite
interesting holographic relation between the Friedmann equation
describing  a radiation-dominated, closed
Friedmann-Robertson-Walker (FRW) universe and Cardy
formula~\cite{Cardy} describing the entropy of matter filling the
universe. The radiation can be represented by a conformal field
theory (CFT) with a large central charge, while the entropy for
the latter can be expressed in terms of the so-called
Cardy-Verlinde formula~\cite{Verl}, a generalized form of the
Cardy formula to any dimension. The Cardy-Verlinde formula was
checked to hold for CFTs with AdS gravity dual, for instance, for
CFTs dual to Schwarzschild-AdS black holes~\cite{Verl}, Kerr-AdS
black holes~\cite{Klemm}, hyperbolic AdS black holes and charged
AdS black holes~\cite{Cai1}, Taub-Bolt AdS instanton
solutions~\cite{Birm}, and Kerr-Newman-AdS black
holes~\cite{Jing}. Verlinde found that the Friedmann equation can
be rewritten in the form of the Cardy-Verlinde formula with the help
of three cosmological entropy bounds, and that when the Hubble
entropy bound is saturated, the Friedmann equation coincides with
the Cardy-Verlinde formula.  This observation is very interesting
in the sense that the Friedmann equation is a dynamic one
describing geometric evolution of the universe while the
Cardy-Verlinde formula is just a formula describing the number of
degrees of freedom of matter in the universe. Therefore, the
Verlinde's observation is indeed an interesting manifestation of
the holographic principle in the cosmological setting.

By considering a moving brane universe in the background of
Schwarzschild-AdS black holes in arbitrary dimensions, the
holographic connection between the geometry and matter can be
realized. A radiation-dominated closed FRW universe appears as an
induced metric on the brane embedded in the bulk background.
Savonije and Verlinde~\cite{SV} interpreted the radiation as the
thermal CFT dual to the bulk Schwarzschild-AdS black hole. Further
they observed that when the brane crosses the bulk black hole
horizon, (i) the entropy and temperature of the universe can be
simply expressed in terms of the Hubble parameter and its time
derivative; (ii) the entropy formula (Cardy-Verlinde formula) of
CFTs in any dimension coincides with the Friedmann equation; and
(iii) the Hubble entropy bound is just saturated by the
Bekenstein-Hawking entropy of bulk black holes.

Since then a lot of studies have been done focusing on the
generalization of \cite{Verl,SV} to various bulk geometries. In
this paper, we are interested in the case where a positive
cosmological constant is present in the universe. Namely we will
discuss the holography for a radiation-dominated closed FRW
universe with a positive cosmological constant. This is motivated
partially by the de Sitter (dS)/CFT correspondence~\cite{Strom},
and partially by recent astronomical observations on supernova
indicating that our universe is accelerating~\cite{Super}, which
can be interpreted as that there might be a positive cosmological
constant in our universe. We will start with a brief review on the
case without the cosmological constant~\cite{Verl} in the next
section, and then discuss the holography in the case with a
positive cosmological constant by introducing a cosmological
D-bound on the entropy of matter in the universe. As an example,
in Sec.~III we investigate the holographic connection between the
Friedmann equation and the Cardy-Verlinde formula in the brane
cosmology in the background of Schwarzschild-de Sitter black
holes. We end this paper in Sec.~IV with some conclusions and
discussions.

\section{Holography in a radiation-dominated closed universe
with a positive cosmological constant}

Let us consider an $(n+1)$-dimensional closed FRW universe
\begin{equation}
\label{2eq1} ds^2 =-d\tau^2 +R^2 d\Omega^2_n,
\end{equation}
where $R$ is the  scale factor of the universe and $d\Omega^2_n$
denotes the line element of an $n$-dimensional unit sphere. The
evolution of the universe is determined by the FRW equations
\begin{eqnarray}
\label{2eq2}
 && H^2 =\frac{16\pi G_{n+1}}{n(n-1)}\frac{E}{V}
-\frac{1}{R^2}
     +\frac{1}{l^2_{n+1}}, \nonumber \\
&& \dot H =-\frac{8\pi G_{n+1}}{n-1}\left (\frac{E}{V} +p\right)
    +\frac{1}{R^2},
\end{eqnarray}
where $H$ represents the Hubble parameter with the definition
$H=\dot R/R$ and the overdot stands for  derivative with respect
to the cosmic time $\tau$,  $E$ is the energy of matter filling
the universe, $p$ is the pressure, $V$ is the volume of the
universe, $V=R^n \Omega_n$ with $\Omega_n$ being the volume of an
$n$-dimensional unit sphere, and $G_{n+1}$ is the Newton constant
in ($n+1$) dimensions. In addition, $l^2_{n+1}$ is related to the
cosmological constant $\Lambda_{n+1}= n(n-1)/2l^2_{n+1}$ in
($n+1$) dimensions.


\subsection{The case without cosmological constant}

In \cite{Verl} Verlinde introduced three entropy
bounds\footnote{In \cite{Verl} the first bound is called the
Bekenstein bound. In fact this bound is slightly different from
the original Bekenstein bound~\cite{Beke} by a numerical factor
$1/n$. So we call this the Bekenstein-Verlinde bound. This bound
could be viewed as the counterpart of the Bekenstein bound in the
cosmological setting~\cite{CMO}.} :
\begin{eqnarray}
\label{2eq3}
 {\rm Bekenstein-Verlinde\ entropy}:&& S_{\rm BV}=\frac{2\pi}{n}ER
   \nonumber \\
 {\rm Bekenstein-Hawking\ entropy}:&& S_{\rm BH}=(n-1)\frac{V}{4G_{n+1}R}
    \nonumber \\
  {\rm Hubble\ entropy}:&& S_{\rm H}=(n-1)\frac{HV}{4G_{n+1}}.
\end{eqnarray}
The Bekenstein-Verlinde entropy bound is supposed to hold for a
weakly self-gravitating universe ($HR \le 1$), while the Hubble
entropy bound works when the universe is in the strongly
self-gravitating phase ($HR \ge 1$). In the case without the
cosmological constant, the Friedmann equation [the first equation
in (\ref{2eq2})] can be rewritten as
\begin{equation}
\label{2eq4} S_{\rm H}=\sqrt{S_{\rm BH}(2S_{\rm BV}-S_{\rm BH})},
\end{equation}
in terms of those three entropy bounds. Expression (\ref{2eq4}) is
similar to the Cardy formula~\cite{Cardy}, an entropy formula of
CFTs in two dimensions. It is interesting to note that when
$HR=1$, one has $S_{\rm BV}= S_{\rm BH}=S_{\rm H}$.

Let us  define a quantity $E_{\rm BH}$ which corresponds to energy
needed to form a black hole with size of the whole universe
through the relation: $ S_{\rm BH}=(n-1)V/4G_{n+1}R \equiv 2\pi
E_{\rm BH} R/n $. With this quantity, the Friedmann equation
(\ref{2eq4}) can be further cast to
\begin{equation}
\label{2eq5}
 S_{\rm H}=\frac{2\pi R}{n}\sqrt{E_{\rm BH}(2E-E_{\rm BH})},
\end{equation}
which takes the same form as the Cardy-Verlinde
formula~\cite{Verl}
\begin{equation}
\label{2eq6}
  S=\frac{2\pi R}{n}\sqrt{E_c(2E-E_c)}.
\end{equation}
This formula is supposed to describe the entropy $S$ of a CFT
living on an $n$-sphere of radius $R$. Here $E$ is the total
energy of the CFT and $E_c$ stands for the Casimir energy of the
system, the non-extensive part of the total energy. Now we suppose
that the entropy of the radiation matter in the FRW universe can
be described by the Cardy-Verlinde formula. Comparing (\ref{2eq5})
with (\ref{2eq6}), one can easily see that if $E_{\rm BH}=E_c$,
$S_{\rm H}$ and $S$ must be equal. In other words, the Hubble
entropy bound is saturated by the entropy of radiation matter in
the universe if the Casimir energy $E_c$ is just enough to form a
black hole with the size of the universe. At that moment,
equations (\ref{2eq5}) and (\ref{2eq6}) coincide with each other.
This implies that the Friedmann equation somehow knows the entropy
formula of radiation-matter filling the universe~\cite{Verl}.
Considering a  brane universe in the background of
Schwarzschild-AdS black holes, Savonije and Verlinde~\cite{SV}
found that when the brane crosses the black hole horizon, the
Hubble entropy bound is saturated by the entropy of black holes in
the bulk.

In Eq.~(\ref{2eq6}) the Casimir energy $E_c$ is defined
as~\cite{Verl}
\begin{equation}
\label{2eq7}
 E_c=n(E +pV -TS),
 \end{equation}
 where $T$ stands for the temperature of the thermal CFT, $p$ is the pressure
 and $V$ is the volume of the system. Further
 Verlinde found that except for the similarity between the
 Friedmann equation (\ref{2eq5}) and the Cardy-Verlinde formula
 (\ref{2eq6}), there is also a similarity between the second
 equation in (\ref{2eq2}) concerning the time derivative of Hubble
 parameter and the equation (\ref{2eq7}) about the Casimir energy
 of CFTs. Let us define a (limiting) temperature
 \begin{equation}
 \label{2eq8}
 T_{\rm H}=-\frac{\dot H}{2\pi H}.
 \end{equation}
 Here the minus sign is necessary to get a positive result. In
 addition, it is assumed that we are in the strongly
 self-gravitating phase with $HR \ge 1$ so that $H\ne 0$ and
 $T_{\rm H}$ is well-defined.
  With this temperature, the second
 equation in (\ref{2eq2}) can be rewritten as
 \begin{equation}
 \label{2eq9}
 E_{\rm BH}=n(E+pV -T_{\rm H}S_{\rm H}).
 \end{equation}
 Thus we see that when the Hubble bound $S_{\rm H}$ is saturated by the
 matter entropy $S$, the (limiting) temperature $T_{\rm H}$ equals
  the thermodynamic temperature $T$ of matter filling the
 universe. Note that like the Hubble entropy bound,  the (limiting) temperature
 $T_{\rm H}$ is a geometric quantity determined by the Hubble parameter
and its time derivative. Clearly the temperature $T_{\rm H}$ is
the minimal one in the strongly self-gravitating
phase~\cite{Verl}.

\subsection{The case with a positive cosmological constant}

Now we turn to the case with a non-vanishing cosmological
constant, and generalize those interesting observations to a
radiation-dominated closed FRW universe with a positive
cosmological constant. We will argue that three cosmological
entropy bounds in (\ref{2eq3}) remain the same forms even when a
cosmological constant is present.

To write down a formula like (\ref{2eq4}), let us first discuss
three entropy bounds in (\ref{2eq3}). The Bekenstein-Verlinde
bound $S_{\rm BV}$ is the counterpart of the Bekenstein entropy
bound~\cite{Beke} in the cosmological setting~\cite{CMO}. It is
supposed to hold for a system with limited self-gravity, which
means that the gravitational self-energy of the system is small
compared to the total energy $E$. Namely, the gravitational effect
on the bound can be neglected. Therefore this bound is independent
of gravity theories. It is also independent of whether or not the
gravity theory under consideration includes  a cosmological
constant. In other words, the form of the Bekenstein-Verlinde
bound should remain unchanged in any gravitational
theory\footnote{ In ~\cite{CM} we show that the Bekenstein entropy
bound always has the form $S_{\rm B}=2\pi E R$ independent of
gravity theories by applying a Geroch process to an arbitrary
black hole.}. Hence even when a positive cosmological constant is
present, the Bekenstein-Verlinde bound still takes the form in
(\ref{2eq3}). As a result, for a radiation-dominated FRW universe
with a cosmological constant, the Bekenstein-Verlinde bound is a
constant because of $E\sim R^{-1}$. Thus once this bound is
satisfied at one time,
 it will be always satisfied at all times if the entropy
 $S$ of matter does not change.

 As for the Bekenstein-Hawking bound in (\ref{2eq3}), it can be
 viewed as the holographic Bekenstein-Hawking entropy of a black
 hole with the size of the universe~\cite{Verl}. Indeed, it varies
 like an area instead of the volume. And for a closed universe it
 is the closest one that can lead to the usual area formula of black
 hole entropy $A/4G$. We know from the thermodynamics of black
 holes that in Einstein gravity  the entropy of a black hole
 is always proportional to its horizon area in spite of whether or not the
 gravitational theory includes  a cosmological
 constant~\cite{TM}. Further, as argued by Verlinde~\cite{Verl},
 the role of $S_{\rm BH}$ is not to serve as a bound on the total
 entropy, but rather on a sub-extensive component of the entropy
 that is associated with the Casimir energy of CFTs. The above
 leads to the conclusion that the Bekenstein-Hawking bound should
 remain unchanged in its form and implication as in the case without
 the cosmological constant.

 Finally we consider the Hubble entropy bound, which is an entropy
 bound for matter in a strongly self-gravitating universe $(HR\ge1$). In such
 a strongly self-gravitating universe,  black holes might occur.
 As argued in \cite{FS,Hubb}, the maximal
 entropy inside the universe is produced by black holes with
 size of the Hubble horizon. The usual holographic argument shows
 that the total entropy should be less than or equal to the
 Bekenstein-Hawking entropy of a Hubble-horizon-sized black hole
 times the number of Hubble regions in the universe. In
 \cite{Verl}, by using a local holographic bound due to
 Fischler and Susskind~\cite{FS} and Bousso~\cite{Bous}, see
 also \cite{Wald}, Verlinde ``derived" the Hubble entropy bound
 in (\ref{2eq3}). It is worth noting that in the ``derivation" of the Hubble
bound, Verlinde used mainly the idea that the entropy flow $S$
through a contracting light sheet is less than or equal to $A/4G$,
where $A$ is the area of the surface from which the light sheet
originates. Hence we insist that the cosmological constant will
not affect the form of the Hubble bound (see also \cite{CM}). This
conclusion is based on the fact that even if a cosmological
constant is present, it will not occur explicitly in the
``derivation" of the Hubble entropy bound.

We conclude that in a closed FRW universe with a cosmological
constant, three bounds introduced in (\ref{2eq3}) are still
applicable.  That is, their forms and implications remain
unchanged even if  the cosmological constant is present. However,
we see that the cosmological constant indeed affects the evolution
of the universe. Is there a similar relation between the Friedmann
equation in (\ref{2eq2}) and the Cardy-Verlinde formula
(\ref{2eq6}) as in the case without the cosmological constant?
 Our key observation is that the positive  cosmological constant provides an
additional entropy measure. When the cosmological constant occurs,
not the Hubble bound, but a new cosmological D-bound plays the
role as the Hubble bound does in the case without the cosmological
constant.

Let us go to the details. We know that in a pure de Sitter
universe, there is a cosmological horizon for an inertial
observer. Like a black hole horizon, the cosmological horizon has
a Hawking temperature and an associated entropy~\cite{GH}. The
entropy is proportional to the area of the cosmological horizon.
 It is a geometric quantity although it has a
statistical origin in quantum gravity. In an asymptotically de
Sitter space, the cosmological horizon shrinks. Applying the
generalized second law of thermodynamics~\cite{Bekenst} to the
cosmological horizon, one can immediately find that the entropy of
matter in the asymptotically de Sitter space has to be bounded
above by the difference (D) between the entropy of the pure de
Sitter space and the Bekenstein-Hawking entropy of the
asymptotically de Sitter space:
\begin{equation}
\label{2eq10}
  S_{\rm m} \le \frac{1}{4 G}\left( A_0-A\right),
\end{equation}
where $A_0$ and $A$ are areas of cosmological horizons for the
pure de Sitter and asymptotically de Sitter spaces, respectively.
This is the so-called D-bound proposed by Bousso in \cite{Bousso}.
The D-bound is closely related to the Bekenstein bound which
applies in flat backgrounds~\cite{Bousso,CMO}.

In our present context, the occurrence of the cosmological
constant does not guarantee that the universe approaches to a de
Sitter phase. When the matter is dominated, the universe behaves
as the case without the cosmological constant: the universe starts
from a big bang, reaches a maximal radius and then re-collapses
with a big crunch. From (\ref{2eq2}), however, we see that for an
empty flat universe\footnote{In that case, $E=p=0$, and the term
$1/R^2$ will be also absent.}, the Hubble radius is just the
cosmological horizon size $l_{n+1}$ of de Sitter space. It implies
that the cosmological constant provides a new entropy measure in
the universe. By analogy with the Hubble entropy
bound~\cite{FS,Hubb,Verl}, we define a quantity
\begin{equation}
\label{2eq11} S_{\rm \Lambda} =(n-1)\frac{V}{4G_{n+1}l_{n+1}}.
\end{equation}
which  is the entropy of a de Sitter horizon times the number of
the regions with the size of the de Sitter horizon  in the
universe. Like the Hubble entropy bound, it is also a geometric
quantity.  Together with the  three entropy bounds in
(\ref{2eq3}),  the Friedmann equation in (\ref{2eq2}) can be
rewritten as
\begin{equation}
\label{2eq12}
  S_{\rm H}^2-S_{\rm \Lambda}^2=S_{\rm BH}(2S_{\rm BV}-S_{\rm
  BH}).
\end{equation}
Further we note that the cosmological horizon in the
asymptotically de Sitter spaces is always less that of
corresponding de Sitter spaces, but one can see from (\ref{2eq2})
that the Hubble radius $H^{-1}$ is not always less than the
cosmological horizon $l_{n+1}$ of de Sitter spaces. As a result,
the left-hand side of equation (\ref{2eq12}) is not always
positive. In addition, we stress that in the case without the
cosmological constant, the Hubble bound in (\ref{2eq5}) is a
geometric quantity, which gives an entropy bound of matter in the
universe when the universe is in the strongly self-gravitating
phase. Considering the D-bound (\ref{2eq10}) of matter in de
Sitter spaces and the similarity between (\ref{2eq4}) and
(\ref{2eq12}), we can define a cosmological entropy bound in the
universe with a positive cosmological constant\footnote{Since a
black hole larger than the cosmological horizon cannot form, one
therefore should have $S_{\rm H} \ge S_{\rm \Lambda}$. As a
result, if $S_{\rm H} <S_{\rm \Lambda}$, a cosmological
singularity might occur during the evolution of the universe.}
\begin{equation}
\label{2eq13}
 S_{\rm D}=\sqrt{|S_{\rm H}^2-S_{\rm
 \Lambda}^2|}.
\end{equation}
We call it the cosmological D-bound, which can be viewed as the
counterpart of the D-bound in the cosmology setting. Note that the
cosmological D-bound is a square root of the difference between
two geometric quantity squares, while the D-bound in de Sitter
spaces is the difference between two geometric quantities.
Furthermore, we would like to mention here that the implications
of these two entropy bounds are quite different.  The D-bound
(\ref{2eq10}) is an entropy bound where the entropy is measurable
to an inertial observer in the de Sitter space(recall the inertial
observer can see only a {\it part} of the de Sitter
space~\cite{GH}), while the cosmological D-bound (\ref{2eq13}) is
supposed to be an entropy bound of matter filling the {\it whole
universe} with a positive cosmological constant when the universe
is in the strongly self-gravitating phase. Therefore, more
precisely, the cosmological D-bound should be regarded as
 the
counterpart of the Hubble entropy bound $S_{\rm H}$ in the case
with a positive cosmological constant. This can be seen clearly
after comparing (\ref{2eq4}) and (\ref{2eq12}). With the choice
(\ref{2eq13}), the Friedmann equation can be cast to the
Cardy-Verlinde form even when the cosmological constant is
present.

On the analogy of the (limiting) temperature $T_{\rm H}$ in
(\ref{2eq8}), we  further  define a new geometric temperature in
our case
\begin{equation}
\label{2eq14}
 T_{\rm D}= -\frac{\dot H}{2\pi
  \sqrt{|1/l_{n+1}^2-H^2|}}.
\end{equation}
Note that this is also a geometric quantity like $T_{\rm H}$ for
the case without the cosmological constant.  With this,  the
second equation in (\ref{2eq2}) can be expressed as
\begin{equation}
\label{2eq15}
 E_{\rm BH}=n(E +pV -T_{\rm D}S_{\rm D}).
\end{equation}
Here the definition of $E_{\rm BH}$ is the same as the one in
(\ref{2eq5}). Like the limiting temperature $T_{\rm H}$, the
geometric temperature $T_{\rm D}$ will equal the thermodynamic
temperature of matter filling the universe when the cosmological
D-bound gets saturated. That is, $T_{\rm D}$ is the lower bound of
thermodynamic temperature during the universe is in the strongly
self-gravitating phase.

Now we turn to the Cardy-Verlinde formula (\ref{2eq6}). In the
form (\ref{2eq6}) it is implicitly assumed that one has $2E-E_c
\ge 0$ for any CFT. In fact, in some circumstances, this condition
does not get satisfied. For example, in the CFT description dual
to the thermodynamics of Schwarzschild-de Sitter black hole
horizon, this quantity is negative~\cite{Cai2}. In that case, the
Cardy-Verlinde formula should be changed to
\begin{equation}
\label{2eq16} S=\frac{2\pi R}{n}\sqrt{E_c(E_c-2E)},
\end{equation}
where the definition of $E_c$ is still the same as the one
(\ref{2eq7}).

In summary, for a radiation-dominated closed FRW universe with a
positive cosmological constant the dynamic equations can be
rewritten as
\begin{eqnarray}
\label{2eq17}
 && S_{\rm D}=\frac{2\pi R}{n}\sqrt{E_{\rm
BH}(2E-E_{\rm
BH})}, \nonumber \\
&& E_{\rm BH}=n(E+pV -T_{\rm D} S_{\rm D}),
\end{eqnarray}
when $S_{\rm H} \ge S_{\Lambda}$, while the entropy of the
radiation can be expressed as
\begin{eqnarray}
\label{2eq18} && S =\frac{2\pi R}{n}\sqrt{E_c(2E-E_c)},
  \nonumber \\
&& E_c=n(E+pV -T S).
\end{eqnarray}
On the other hand, when $S_{\rm H} \le S_{\rm \Lambda}$, the
dynamic equations can be rewritten as
\begin{eqnarray}
\label{2eq19}
 && S_{\rm D}=\frac{2\pi R}{n}\sqrt{E_{\rm
BH}(E_{\rm
BH}-2E )}, \nonumber \\
&& E_{\rm BH}=n(E+pV -T_{\rm D} S_{\rm D}),
\end{eqnarray}
and the entropy expressions are
\begin{eqnarray}
\label{2eq20}
 && S =\frac{2\pi R}{n}\sqrt{E_c(E_c-2E)},
  \nonumber \\
&& E_c=n(E+pV -T S).
\end{eqnarray}
When the cosmological D-bound is saturated by the entropy $S$ of
radiation matter, both sets of equations (\ref{2eq17}) [or
(\ref{2eq19})] and (\ref{2eq18}) [or (\ref{2eq20})] coincide with
each other, just like the case without the cosmological constant.
The cosmological D-bound in (\ref{2eq13}) provides an entropy
bound for matter filling the universe when the universe is in the
strongly self-gravitating phase\footnote{When the universe is in
the weakly self-gravitating phase, as argued in the above, the
Bekenstein-Verlinde bound still works well.}. Namely, the
cosmological D-bound plays the same role as the Hubble bound does
in the case without cosmological constant.

\section{Brane cosmology in the background of Schwarzschild-de
Sitter black holes}

\subsection{Thermodynamics of Schwarzschild-de Sitter black holes}

Consider an ($n+2$)-dimensional Schwarzschild-de Sitter black
hole, whose line element is
\begin{equation}
\label{3eq1} ds^2 =- f(r) dt^2 +f(r)^{-1}dr^2 +r^2 d\Omega_{n}^2.
\end{equation}
Here
$$ f(r) =1 -\frac{\omega_n M}{r^{n-1}} -\frac{r^2}{l_{n+2}^2}, \ \ \
 \omega_n=\frac{16\pi G_{n+2}}{n \Omega_n},$$
$M$ stands for the mass of the Schwarzschild-de Sitter black hole
in the definition of  Abbott and Deser~\cite{AD}, $G_{n+2}$
denotes the $(n+2)$-dimensional Newton constant, and $l_{n+2}$
represents the cosmological radius of the ($n+2$)-dimensional de
Sitter universe. When $M=0$, the solution (\ref{3eq1}) reduces to
a  de Sitter space with a cosmological horizon at $r_c=l_{n+2}$.
When $M$ increases from $M=0$, a black hole horizon appears and
grows, while the cosmological horizon shrinks. Finally the black
hole horizon $r_{\rm BH}$ touches the cosmological horizon $r_{\rm
CH}$ when
$$ M =M_N \equiv \frac{2}{\omega_n (n+1)}\left
(\frac{n-1}{n+1}l^2_{n+2}\right)^{(n-1)/2}.$$
 This is the Nariai black hole, the maximal black hole in de
 Sitter space. When $M >M_N$, both the two horizons disappear and
 the solution describes a naked singularity. When $M <M_N$, the
 equation $f(r)=0$ has two real roots, the larger one is the
 cosmological horizon, while the smaller one is the black hole
 horizon.

The Hawking temperature $T_{\rm HK}$ and entropy $S$ associated
with the
  black hole horizon are~\cite{Cai2}
  \begin{equation}
  \label{3eq2}
  T_{\rm HK} =\frac{1}{4\pi
  r_{\rm BH}}\left((n-1)-(n+1)\frac{r_{\rm BH}^2}{l^2_{n+2}}\right), \ \ \ \
 S =\frac{r_{\rm BH}^n \Omega_n}{4G_{n+2}}.
\end{equation}
With the identification $E=M$ and the definition
$E_c=(n+1)E-nT_{\rm HK}S~$\footnote{Here it is assumed that the
thermodynamics of the Schwarzschild-de Sitter black hole can be
described in terms of a CFT. Furthermore, $E_c$ here is obtained
from the definition (\ref{2eq7}) with the equation of state $
p=E/nV $ of CFT's.}, one can easily obtain
\begin{equation}
 E_c =\frac{2nr_{\rm BH}^{n-1}\Omega_n}{16\pi G_{n+2}}, \ \ \
 2E-E_c =-\frac{2n r_{\rm BH}^{n+1} \Omega_n}{16\pi G_{n+2}l^2_{n+2}}.
 \end{equation}
Clearly the entropy (\ref{3eq2}) can be expressed by the
Cardy-Verlinde formula~\cite{Cai2}
\begin{equation}
S =\frac{2\pi l_{n+2}}{n} \sqrt{E_c (E_c-2E)}.
\end{equation}
If one rescales the energies by a factor $R/l_{n+2}$, the above
equation is changed to
\begin{equation}
\label{3eq5} S =\frac{2\pi R}{n}\sqrt{E_c(E_c-2E)}.
\end{equation}
Note that this expression is  exactly the same as the entropy
formula in (\ref{2eq20}).

\subsection{Brane dynamics in the background of Schwarzschild-de
Sitter black holes}

Let us introduce an $(n+1)$-dimensional brane with tension $\sigma
$ moving in the background of the Schwarzschild-de Sitter black
holes (\ref{3eq1}). Its dynamics is determined by the following
action~\cite{SV,Wall}
\begin{equation}
\label{3eq6} S_{\rm brane}=\frac{1}{8\pi G_{n+2}}\int_{\partial
M}d^{n+1}x \sqrt{-h}K +\frac{1}{8\pi G_{n+2}}\int_{\partial
M}d^{n+1}x\sqrt{-h} \sigma.
\end{equation}
Here the brane is viewed as boundary of the bulk spacetime
(\ref{3eq1}), $K$ is the extrinsic curvature for the boundary with
the induced metric $h_{ab}$. Using the Israel junction
equation~\cite{Israel}, one has the equation of motion of the
brane
\begin{equation}
\label{3eq7}
 K_{ab}=\frac{\sigma}{n}h_{ab}.
\end{equation}
The brane cosmology in the Schwarzschild-de Sitter black holes has
been first considered in \cite{Ogush}. The holography in brane
cosmology in various asymptotically de Sitter spaces has  also
been discussed in~\cite{Ogush,deSitter}\footnote{However, our
philosophy in understanding the holography in the case with a
cosmological constant is different from those in
\cite{Ogush,deSitter}. We will discuss this point at the end of
this paper.}.
 Now let us specify the location
of the brane as $r=r(t)$. We introduce a cosmic time $\tau$ so
that $t=t(\tau)$ and $r=r(\tau)$ and require
\begin{equation}
\label{3eq8}
 f(r)\left(\frac{dt}{d\tau}\right)^2
 -\frac{1}{f(r)}\left(\frac{dr}{d\tau}\right)^2 =1,
 \end{equation}
 which implies that the brane moves along a radial time-like geodesic
in the background (\ref{3eq1})\footnote{The dynamics of brane
along a radial space-like geodesic in various asymptotically de
Sitter backgrounds has also been discussed in
\cite{Ogush,deSitter}.}. In that case, the induced metric $h_{ab}$
on the brane becomes
\begin{equation}
\label{3eq9} ds^2 =-d\tau^2 +R^2(\tau)d\Omega_n^2,
\end{equation}
which is just an ($n+1)$-dimensional closed FRW universe metric
(\ref{2eq1}) with  scale factor $R(\tau)=r(\tau)$.

Calculating the extrinsic curvature for the brane and then from
the equation (\ref{3eq7}), we have
\begin{equation}
\label{3eq10}
 \frac{dt}{d\tau}= \frac{\sigma R}{nf(R)}.
\end{equation}
Substituting into (\ref{3eq8}) yields
\begin{equation}
\label{3eq11} H^2 = \frac{\omega_n M}{R^{n+1}}
 -\frac{1}{R^2}+\frac{1}{l_{n+2}^2} +\frac{\sigma^2}{n^2},
 \end{equation}
 The time derivative of the Hubble parameter is
 \begin{equation}
 \label{3eq12}
\dot H =-\frac{(n+1)\omega_n M}{2R^{n+1}} +\frac{1}{R^2}.
\end{equation}
These equations describe a radiation-dominated closed FRW universe
with a positive cosmological constant $\Lambda_{n+1}=
n(n-1)/2l^2_{n+1}$ with
\begin{equation}
 \label{3eq13}
 \frac{1}{l_{n+1}^2}=
    \frac{1}{l_{n+2}^2} +
       \frac{\sigma^2}{n^2},
 \end{equation}
 from which we see $l^2_{n+1} < l^2_{n+2}$. Now we consider the solution
  of (\ref{3eq11}). As an example, let us discuss the special case of
$n=3$. The generalization to other dimensions is straightforward.
In that case, the solution has been found for three different
cases depending on the parameter $\omega_4 M/l^2_{n+1}$ in
\cite{PS}, where the authors discussed the dynamics of  a
non-critical
 brane in the Schwarzschild-AdS black hole. Defining $x=R^2$, we
 can rewrite (\ref{3eq11}) as
 \begin{equation}
 \label{3eq14}
 \dot x^2 =\frac{4}{l_4^2}(x-x_+)(x-x_-),
 \end{equation}
where
\begin{equation}
x_{\pm} =\frac{l^2_4}{2}\left(1 \pm
\sqrt{1-4\omega_4M/l_4^2}\right).
\end{equation}
Note that when $x=x_{\pm}$, one has $H=0$. Actually, $x_{\pm}$ are
turning points of the brane.

(1) When $4\omega_4 M =l^2_4$, one has $x_+=x_-$.  In this case,
the brane has only one turning point. However, the solution has
two branches:
\begin{itemize}
\item  $x \in (0, x_+]$. In this case the solution is
\begin{equation}
   R^2(\tau) = \frac{l_4^2}{2}\left(1-e^{-2\tau/l_4}\right),
     \ \ \  \tau \in [0, \infty).
\end{equation}
Accordingly the Hubble parameter reads
\begin{equation}
H= \frac{1}{l_4}\frac{e^{-2\tau/l_4}}{1-e^{-2\tau/l_4}}.
\end{equation}
In this branch, $ 0\le H < \infty$.  Since  the black hole horizon
$x_{\rm BH}$ falls in the range of $0 <x_{\rm BH} < x_+$,  we find
$ H <1/l_4$ in the range of $ x_{\rm BH} < x < x_+$.
 The brane trajectory is plotted in Fig.~1 as the curve (I). The brane
   universe starts from a big bang, reaches a maximal radius
   and then re-collapses to a big crunch.

\item  $ x \in [x_+,\infty)$. In this case, the solution is given by
   \begin{equation}
   \label{3eq16}
   R^2 =x(\tau)= \frac{l^2_4}{2}\left(1 +e^{2\tau/l_4}\right), \ \
    \ \tau \in (-\infty, \infty).
   \end{equation}
   And the Hubble parameter takes the expression
   \begin{equation}
   H =\frac{1}{l_4}\frac{e^{2\tau/l_4}}{1+e^{2\tau/l_4}}.
   \end{equation}
   Clearly in this case one has $0< H < 1/l_4$. The brane
trajectory is plotted in Fig.~1 as the curve (II). The universe
contracts with a big radius, reaches a minimal radius (the minimal
radius is equal to the maximal radius in the previous branch), and
then bounces to infinity. This is a singularity-free cosmology.

\end{itemize}

(2) When $4\omega_4 M > l_4^2$, the brane has no turning point.
Namely, there is no point which has $H=0$ along the geodesic of
the brane. In this case one has the solution
    \begin{equation}
   \label{3eq20}
    R^2(\tau)= \frac{l_4^2}{2}\left (1 +\sqrt{4\omega_4M/l_4^2
    -1}\sinh( 2\tau /l_4)\right),
    \end{equation}
    where $\tau_0 \le \tau < \infty$ with
    $$ \sinh(2\tau_0/l_4) =-\left(4\omega_4 M/l_4^2-1
      \right)^{-1/2}.$$
   The evolution of the Hubble parameter is given by
   \begin{equation}
   H
   =\frac{1}{l_4}\frac{\sqrt{4\omega_4M/l_4^2-1}\cosh(2\tau/l_4)}
    {1+\sqrt{4\omega_4M/l_4-1}\sinh(2\tau/l_4)}.
    \end{equation}
 We find that outside the black hole horizon, $H<1/l_4$. Actually, there
exists another solution for the equation (\ref{3eq14}). But this
solution describes the same movement of the brane as the solution
(\ref{3eq20}) does. So we do not present it here. The trajectory
of the brane universe is plotted in Fig.~2. In this case, the
universe starts with a big bang and then expands forever.

    (3) When $4\omega_4 M <l_4^2$, the brane has two turning
 points $x_{\pm}$. The range $x \in [x_-,x_+]$ is not allowed  since in which
$H^2 \le 0$. As a result,  solution of
  equation (\ref{3eq14}) has   two branches:
  \begin{itemize}
  \item $x\in (0, x_-]$. The solution is
  \begin{equation}
  R^2(\tau)= \frac{l_4^2}{2}\left(1
  -\sqrt{1-4\omega_4M/l_4^2}\cosh(2\tau/l_4)\right),
  \end{equation}
  where $\tau$ takes value in the range $-\tau_c \le \tau \le
  \tau_c$ with
\begin{equation}
 \cosh(2\tau_c/l_4)=(1-4\omega_4M/l_4)^{-1/2}.
 \end{equation}
  When $\tau =\pm \tau_c$, one has $R=0$ and $H= \infty$.
  The Hubble parameter
  \begin{equation}
  H
  =\frac{1}{l_4}\frac{\sqrt{1-4\omega_4 M/l_4^2}\sinh(2\tau/l_4)}
    {1-\sqrt{1-4\omega_4M/l_4^2}\cosh(2\tau/l_4)},
    \end{equation}
    from which we see that $H=0$ when $\tau=0$.
Note that because of the relation (\ref{3eq13}), it is easy to see
that the black hole horizon $x_{\rm BH} \in
 (0, x_-)$, while the cosmological horizon
$x_{\rm CH} \in (x_+, \infty)$.
 Thus we find that in this branch one has $H <1/l_4$ as
 the brane stays outside the black hole horizon  of the bulk
 background. The brane trajectory is plotted in Fig.~3 as the
 curve (I).

  \item $x\in[x_+,\infty)$. In this branch the solution is
  \begin{equation}
  R^2(\tau)=\frac{l^2_4}{2}\left( 1+\sqrt{1-4\omega_4
  M/l_4^2}\cosh(2\tau/l_4)\right), \ \ \ \tau \in (-\infty,
  \infty).
  \end{equation}
  And the Hubble parameter is given by
  \begin{equation}
  H =\frac{1}{l_4}\frac{\sqrt{1-4\omega_4 M/l_4^2}\sinh(2\tau/l_4)}
    {1+\sqrt{1-4\omega_4M/l_4^2}\cosh(2\tau/l_4)}.
    \end{equation}
  At $\tau=0$, one has $H=0$. The trajectory of the brane is plotted
  as the curve (II) in Fig.~3.
\end{itemize}
We conclude that the evolution of the brane depends on value of
the parameter $4\omega_4M/l_4^2$ and its initial position. Since
we are interested in a radiation-dominated universe beginning with
a big bang, so the solutions in the branch $x \in (0,x_+]$ of case
(1) and in the branch $x \in (0, x_-]$  of case (3) are suitable,
respectively, for our purpose. Inspecting them, we find that $H
<1/l_4$ always holds when the brane stays outside the bulk black
hole horizon. Further we mention that in the above discussions,
the condition $4\omega_4M <l^2_5$ is assumed to hold, which
implies that the black hole horizon is always present.

\subsection{Holography in the brane cosmology}

In the brane world scenario with an AdS bulk, the tension of the
brane can be adjusted to result in a so-called critical brane on
which the effective cosmological constant vanishes~\cite{SV}. In
the present case, one can see from (\ref{3eq13}) that it is
impossible to obtain a vanishing cosmological constant on the
brane. Now we set~\footnote{For the case $\sigma \ne n/l_{n+2}$,
we have a discussion in the following section. Also in that case,
the relations (\ref{3eq28}) and (\ref{3eq29}) have to be changed.}
\begin{equation}
\label{3eq27}
 \sigma = n/l_{n+2}.
\end{equation}
In that case the Newton constant on the brane has the relation
\begin{equation}
\label{3eq28}
 G_{n+1} = \frac{n-1}{ l_{n+2}}G_{n+2},
\end{equation}
to the Newton constant in the bulk. This relation is the same as
that for a critical brane in AdS bulk~\cite{SV}. Furthermore the
parameter $M$ in the solution (\ref{3eq1}) is the black hole mass
measured in the bulk coordinates~\cite{AD}. According to the
relation (\ref{3eq10}), the holographic energy $E$ measured on the
brane is\footnote{Due to existence of the cosmological horizon in
the bulk, this relation is not justified well as the case for the
AdS bulk~\cite{SV}. Even for the latter case, there exists a
different viewpoint, for example, see \cite{Padi}, which argued
that this relation holds only near the boundary of AdS space.
However, we note that the rescaling (\ref{3eq29}) indeed gives a
scale relation for a radiation matter in universe. Further, the
relation (\ref{3eq29}) holds at least for small black holes as in
the AdS case.} \footnote{In the case of AdS bulk, the boundary of
the AdS space is a Lorentz spacetime, it is natural then to expect
a holographic description for a dynamic brane near the
boundary~\cite{SV}. The boundary of dS space is a Euclidean space,
however,  the brane as the boundary of the bulk spacetime we have
discussed so far is Lorentz, so in the spirit of dS/CFT
correspondence it is very strange that there exists a holographic
description for the dynamic brane in the asymptotically dS space.
We thank the referee for pointing this out to us. We guess that
this might be related to that the horizon thermodynamics has a
universal $2$-dimensional CFT description~\cite{Carlip}. For
example, for asymptotically non-AdS spaces, there is also a
holographic description for a dynamic brane and the resulting
equation of motion can also be cast to the Cardy-Verlinde
form~\cite{Caizhang}. Therefore the holographic connection between
the dynamics of the brane and the bulk horizon thermodynamics
might be beyond the AdS/CFT and dS/CFT correspondences.}
\begin{equation}
\label{3eq29}
 E =\frac{l_{n+2}}{R}M.
\end{equation}
Substituting (\ref{3eq27}), (\ref{3eq28}) and (\ref{3eq29}) into
(\ref{3eq11}) and (\ref{3eq12}), we have
\begin{eqnarray}
\label{3eq30}
 && H^2 =\frac{16\pi G_{n+1}}{n(n-1)}\frac{E}{V}
-\frac{1}{R^2}
     +\frac{1}{l^2_{n+1}}, \nonumber \\
&& \dot H =-\frac{8\pi G_{n+1}}{n-1}\left (\frac{E}{V} +p\right)
    +\frac{1}{R^2},
\end{eqnarray}
with $l^2_{n+1}=l^2_{n+2}/2$ and $p=E/nV$. These two equations are
the same as the ones in (\ref{2eq2}).  The equation of state
$p=E/nV$ is just the one for radiation matter (or more general
CFTs). As a result, the discussions on holography in Sec.~2 are
applicable here.

Suppose the brane moves between the bulk black hole horizon and
cosmological horizon\footnote{Note that the brane will not cross
the cosmological horizon in the case where the radiation is
dominated, see the previous subsection.}. Since the brane is
viewed as the boundary of the bulk spacetime, the entropy of
holographic matter (radiation) on the brane is just the entropy of
black hole horizon, which is a constant during the evolution of
the brane universe. However, the entropy density varies with time
as
\begin{equation}
\label{3eq31}
 s \equiv \frac{S}{V}=\frac{r_{\rm BH}^n}{4G_{n+2}R^n}
    =\frac{(n-1)r_{\rm BH}^n}{4G_{n+1}l_{n+2}R^n},
\end{equation}
and the energy density of radiation-matter
\begin{equation}
\label{3eq32}
 \rho \equiv \frac{E}{V}=\frac{n r_{\rm
BH}^{n-1}l_{n+2}}{16\pi G_{n+2} R^{n+1}} \left (1-\frac{r_{\rm
BH}^2}{l^2_{n+2}}\right),
\end{equation}
in terms of the black hole horizon radius $r_{\rm BH}$. Further
from the scaling relation (\ref{3eq29}), the temperature $T$ on
the brane is given by
\begin{equation}
\label{3eq33}
 T=\frac{l_{n+2}}{R}T_{\rm HK}=\frac{l_{n+2}}{4\pi
  r_{\rm BH}R}\left((n-1)-(n+1)\frac{r_{\rm BH}^2}{l^2_{n+2}}\right).
\end{equation}
Applying the first law of thermodynamics to the radiation matter
in the brane universe, one has
\begin{equation}
Tds = d\rho +n(\rho +p -T s)\frac{dR}{R}.
\end{equation}
Following \cite{SV} and defining
\begin{equation}
\gamma = \frac{n}{2}(\rho +p -Ts)R^2,
\end{equation}
we have
\begin{equation}
\label{3eq36}
 \gamma =\frac{nr_{\rm BH}^{n-1}l_{n+2}}{16\pi
G_{n+2}R^{n-1}}
       =\frac{n(n-1)r_{\rm BH}^{n-1}}{16\pi G_{n+1} R^{n-1}}.
 \end{equation}
Obviously, the non-vanishing $\gamma$ is the effect of spatial
curvature $1/R^2$. That is, because of the spatial curvature, the
energy of thermodynamics system has a non-extensive part.
Otherwise, the quantity $\gamma$ must vanish due to the Euler
relation. From (\ref{3eq32}), one can see that $\gamma$ is just
the first term of the energy density (\ref{3eq32}) divided by
$1/R^2$. Further one can find that the entropy density
(\ref{3eq31}) can be expressed as
\begin{equation}
\label{3eq37}
 s=\frac{4\pi }{n}\sqrt{\gamma
\left(\frac{\gamma}{R^2}-\rho \right)}.
\end{equation}
Now we consider a special moment that the brane crosses the bulk
black hole horizon. On that time, one has $R=r_{\rm BH}$. From
(\ref{3eq11}) we see
\begin{equation}
H^2 =\frac{1}{l_{n+2}^2}.
\end{equation}
Note that we have taken $\sigma^2/n^2 =1/l_{n+2}^2$. At that
moment the cosmological D-bound in (\ref{2eq13}) turns out to be
\begin{equation}
S_{\rm D}= \frac{( n-1)r_{\rm BH}^n\Omega_n}{4G_{n+1}l_{n+2}}
             =\frac{ r_{\rm BH}^n \Omega_n}{4G_{n+2}},
\end{equation}
which is just the black hole horizon entropy (\ref{3eq2}). That
is, when the brane  crosses the bulk black hole horizon, the
cosmological D-bound is saturated by the entropy of the bulk black
hole.  At that time, the geometric temperature in (\ref{2eq14}) is
given by
\begin{equation}
T_{\rm D} =\frac{l_{n+2}}{4\pi r_{\rm BH}^2}\left ((n-1) -(n+1)
\frac{r_{\rm BH}^2}{l_{n+2}^2}\right),
\end{equation}
which equals  the temperature (\ref{3eq33}) of radiation filling
the universe. Thus, we find that when the brane crosses the black
hole horizon,  the FRW equations (\ref{2eq19}) coincide with the
equations (\ref{3eq37}) and (\ref{3eq36}) which describe the
entropy of radiation matter in the brane universe with a positive
cosmological constant. Thus we reach the same conclusion as in the
case without the cosmological constant~\cite{SV}.

\section{Conclusion and discussion}

We  have discussed the holography in a radiation-dominated, closed
FRW universe with a positive cosmological constant. By introducing
the cosmological D-bound (\ref{2eq13}) on the entropy of matter in
the universe, the Friedmann equation describing the evolution of
the universe can be rewritten in the form of Cardy-Verlinde
formula which describes the degrees of freedom of radiation matter
filling the universe. When the cosmological D-bound is saturated
by the entropy of matter, these two equations coincide with each
other. Thus we have successfully generalized interesting
observations by Verlinde~\cite{Verl,SV} on the holographic
connection between the Friedmann equation and Cardy-Verlinde
formula to the case with a positive cosmological constant. By
considering a brane universe in the background of Schwarzschild-de
Sitter black holes, we have found that the cosmological D-bound is
saturated when the brane crosses the black hole horizon in the
bulk background. At that moment, the Friedmann equation and
Cardy-Verlinde formula coincide with each other, and the
introduced geometric temperature $T_{\rm D}$ in (\ref{2eq14})
equals  the thermodynamic temperature $T$ of the radiation matter.

We stress that when discussing the holographic connection in the
brane universe in the Schwarzschild-de Sitter black holes, we have
taken a special tension (\ref{3eq27}), which is the same as in the
case for critical brane in the AdS bulk. Only in that case, the
Friedmann equation and Cardy-Verlinde formula coincide with each
other very well. We point out here that if the brane tension is
arbitrary, the Friedmann equation still has a  form as the
Cardy-Verlinde formula,  but a factor $\sigma l_{n+2}/n$ will
appear in (\ref{3eq37}). Furthermore, for the radiation matter
dual to the black holes in de Sitter spaces, we see from
(\ref{3eq37}) that $\rho - \gamma/R^2 <0$. If
 considering a non-critical brane cosmology in the Schwarzschild-AdS black holes,
 one will see that in that case $\rho-\gamma/R^2>0$. This case corresponds to the
holographic connection described by equations (\ref{2eq17}) and
(\ref{2eq18}). In addition, if the cosmological constant becomes
negative, the minus sign in front of $S_{\rm \Lambda}$ has to be
changed to plus. The quantity (\ref{2eq11}) then will lose its
interpretation, but all formulas will still work well.

We have noticed that the holography was discussed in many
literatures in the case with a cosmological constant, for example,
see \cite{Ogush,deSitter,Padi,Wang,Nojiri,Youm,Medved}. However,
our understanding is different from those in existing literature:
in some papers the cosmological constant term is incorporated to
the Bekenstein-Verlinde entropy bound; in some papers this term is
kept as an independent term. In those literature the Friedmann
equation and the Cardy-Verlinde formula have not a same form, and
when the Hubble bound is saturated, these two formulas do not get
matched.  Finally we point out that at the end of
paper~\cite{Verl}, Verlinde mentioned that when the cosmological
constant does not vanish, the Hubble entropy bound needs to be
modified by replacing $H$ with the square root of
$H^2-1/l_{n+1}^2$. But, as argued in this paper, three entropy
bounds: Bekenstein-Verlinde bound, Bekenstein-Hawking bound and
Hubble bound in (\ref{2eq3}) still have the same forms as the case
without the cosmological constant, even when the cosmological
constant is present. The cosmological D-bound introduced in this
work provides a new entropy bound of matter in the strongly
self-gravitating universe ($HR>1$) with a positive cosmological
constant and makes all formulas work so nicely as the case without
the cosmological constant.

\section*{Acknowledgment}
We thank X.J. Wang for help in drawing the figures. R.G.C. is
grateful to Relativity Research Center and School of Computer
Aided Science, Inje University for warm hospitality during his
visit. The work  of R.G.C. was supported in part by a grant from
Chinese Academy of Sciences and a grant from Ministry of
Education, PRC, and by the Ministry of Science and Technology of
China under grant No. TG1999075401. Y.S.M. acknowledges partial
support from the KOSEF grant, Project Number:
R02-2002-000-00028-0.

\begin{figure}
\psfig{file=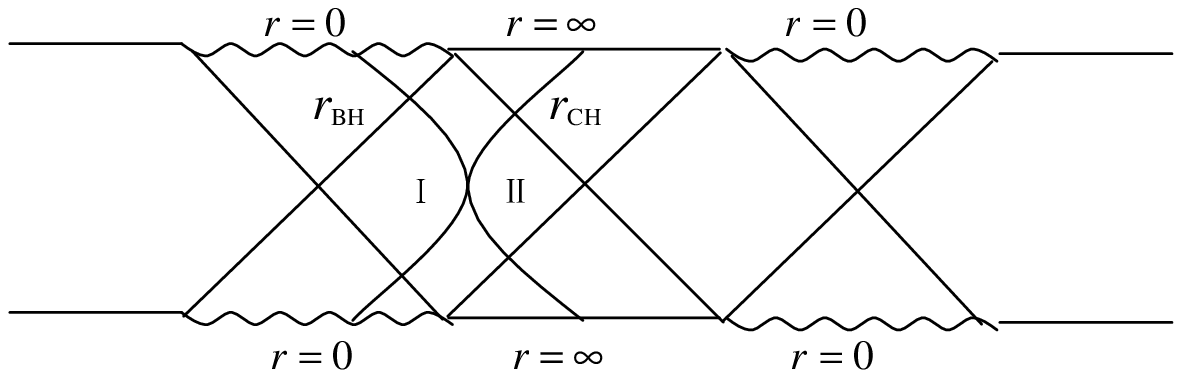,height=55mm,width=120mm,angle=0}
  \caption{The Penrose diagram of the Schwarzschild-de Sitter space. The
  curve (I) describes the trajectory of brane for the branch $x \in (0,x_+]$,
  and the curve (II) for the branch $x\in [x_+, \infty)$.}
 \end{figure}

\begin{figure}
\psfig{file=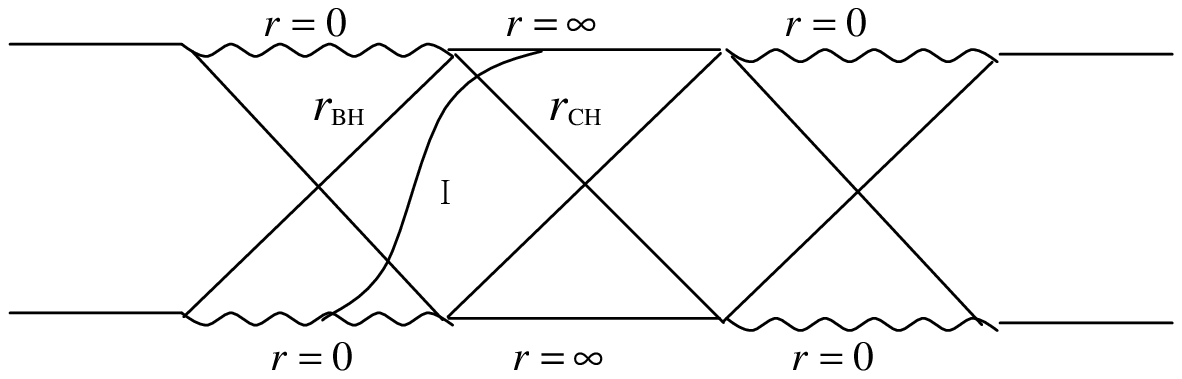,height=55mm,width=120mm,angle=0}
  \caption{The Penrose diagram of the Schwarzschild-de Sitter space. The
  curve (I) describes the trajectory of brane in the case $4\omega_4M<l^2_4$.}
 \end{figure}

 \begin{figure}
\psfig{file=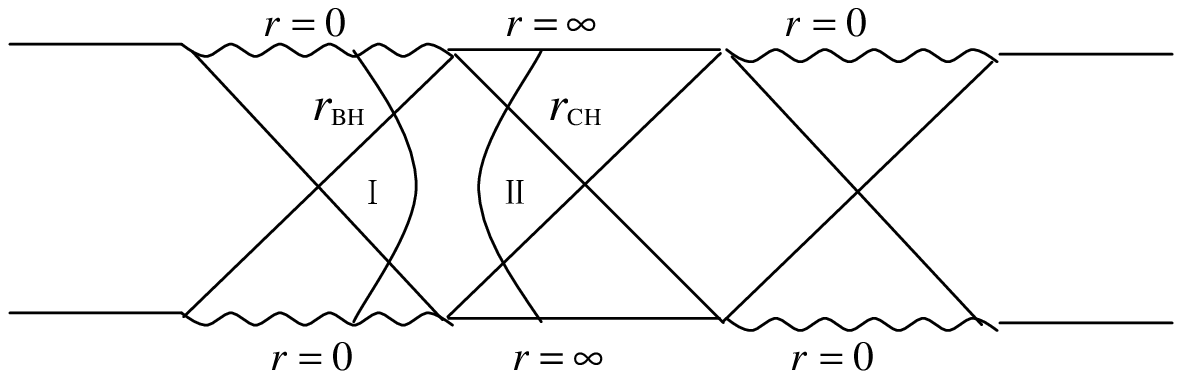,height=55mm,width=120mm,angle=0}
  \caption{The Penrose diagram of the Schwarzschild-de Sitter space. The
  curve (I) describes the trajectory of brane for the branch $x \in (0,x_-]$,
  and the curve (II) for the branch $x\in [x_+, \infty)$.}
 \end{figure}
\end{document}